\documentclass[letterpaper]{article}
\usepackage[preprint]{aaai2027}
\usepackage[hyphens]{url}
\usepackage{graphicx}
\usepackage{natbib}
\usepackage{caption}
\usepackage{algorithm}
\usepackage{algorithmic}

\usepackage{booktabs}

\usepackage{amsmath}

\title{CipherSight: Robust Website Fingerprinting via Record-Resource Semantic Supervision under Distribution Shifts}
\author{
    Runhan Song\textsuperscript{\rm 1,2},
    Qiqi Liu\textsuperscript{\rm 3},
    Chuanzhou Pan\textsuperscript{\rm 2,4}\\
    Zhenquan Ding\textsuperscript{\rm 2},
    Youquan Xian\textsuperscript{\rm 2,4},
    Chongru Fan\textsuperscript{\rm 2,4}\\
    Lei Cui\textsuperscript{\rm 1,2,*},
    Wei Wang\textsuperscript{\rm 2,*},
    Zhiyu Hao\textsuperscript{\rm 2}
}
\affiliations{
    \textsuperscript{\rm 1}Faculty of Computing, Harbin Institute of Technology\\
    \textsuperscript{\rm 2}Zhongguancun Laboratory\\
    \textsuperscript{\rm 3}University of Chinese Academy of Sciences\\
    \textsuperscript{\rm 4}School of Cyberspace Security, Beijing University of Posts and Telecommunications\\
    \textsuperscript{*}Corresponding authors: \texttt{cuilei@zgclab.edu.cn}; \texttt{wangwei@zgclab.edu.cn}
}

\begin{document}

\maketitle

\begin{abstract}

HTTPS website fingerprinting (WF) aims to identify visited websites from metadata observable in encrypted traffic. However, real-world deployments introduce a significant out-of-distribution (OOD) problem caused by temporal and geographic changes, while previously unseen websites are common in open-world scenarios. Existing methods primarily learn from raw TCP packet sequences and struggle to capture stable and generalizable website representations, resulting in performance degradation under practical conditions.

We propose CipherSight, a TLS-record-based hierarchical framework for robust HTTPS WF. Unlike existing approaches that rely on TCP packet sequences and are sensitive to transport-layer artifacts, CipherSight learns website representations from TLS records by jointly encoding multiple record-level attributes. It introduces a hierarchical architecture that captures both intra-flow dependencies among TLS records and inter-flow interactions across concurrent flows, enabling the model to exploit structural patterns in HTTPS traffic. Besides, to learn robust representations, CipherSight employs a masked record modeling (MRM) task to capture contextual traffic semantics and leverages fine-grained record-resource annotations as privileged supervision through structure-aware objectives and semantic distillation.
Experiments show that CipherSight achieves 95.41\% accuracy across more than 2,000 website classes in the closed-world setting and maintains over 90\% accuracy under both temporal and geographic drift, consistently outperforming all evaluated baselines.

\end{abstract}

\section{Introduction}

Website fingerprinting (WF) aims to identify the websites visited by users from metadata observable in encrypted traffic, without decrypting application payloads. It provides a practical approach for studying residual information leakage in encrypted communications and evaluating the effectiveness of traffic analysis defenses~\citep{tor_leakage}.

Early WF studies primarily relied on explicit signals, including DNS queries, TLS Server Name Indication (SNI), and destination IP addresses~\citep{hw}. However, these signals are becoming increasingly unreliable due to emerging privacy-preserving mechanisms, e.g., Encrypted Client Hello (ECH) hides SNI, encrypted DNS protocols protect domain resolution, and shared CDN services weaken the direct mapping between IP addresses and websites~\citep{rfc9849,rfc7858,rfc8484,rfc9250}. Consequently, recent HTTPS WF approaches, inspired by Tor-based traffic analysis, exploit ciphertext-observable side channels, ranging from packet-level features, e.g., lengths, directions, and timing patterns~\citep{df,varcnn,tiktok,ares} to HTTPS-specific flow statistics and higher-layer traffic structures~\citep{adu,hw,stcwf,ma2024flowcontext}.

Although existing methods achieve high performance under closed-world settings, real-world deployment remains challenging mainly due to distribution shifts and unseen traffic in open-world scenarios.
Website content evolves over time, while routing paths, CDN selection, and localized resources vary across regions, causing substantial shifts in encrypted traffic distributions. We therefore evaluate temporal and geographic drift as OOD generalization settings, together with open-world recognition, which requires detecting websites absent from training~\citep{koh2021wilds,yuan2024hswf,jansen2023repositioning}.
There are three challenges that arise in these settings.

\subsection{Challenges}

First, packet-level representations are highly sensitive to transport-layer variability. TCP segmentation, retransmissions, and dynamic network conditions can introduce substantial variations, causing the same website resources to be mapped into substantially different TCP packet sequences. Such network-induced variations introduce noise into learned website representations and hinder robust generalization. According to our analysis, more than 80\% of TCP packet sequences are inconsistent across repeated traffic captures, highlighting their instability under diverse real-world network conditions.

Second, modern HTTPS page loads involve multiple concurrent flows. Tor-oriented methods typically flatten a trace into a single packet sequence~\citep{df,varcnn,tiktok,ares}, implicitly overlooking the inherent multi-flow structure of HTTPS traffic. Recent HTTPS-specific approaches have started to incorporate flow context~\citep{stcwf,ma2024flowcontext}. However, simply modeling individual flows remains insufficient. Packets within the same flow share connection-level and transport semantics, while interactions among concurrent flows collectively reveal the organization of webpage loading. Thus, a robust representation should capture these two complementary levels of dependencies.

Third, encryption creates a semantic gap between observable traffic patterns and underlying webpage composition. Web resources are serialized into HTTP messages, encrypted into TLS records, and transported as TCP packets. While resource-level properties, such as type, size, and request semantics, more directly characterize webpage composition, encrypted traces expose only indirect manifestations of these properties. Training solely with website-level labels leaves the correspondence between TLS records and resource semantics largely unexplored, preventing the model from leveraging fine-grained resource information for robust WF.

\begin{figure}[ht]
    \centering
    \includegraphics[width=1\linewidth]{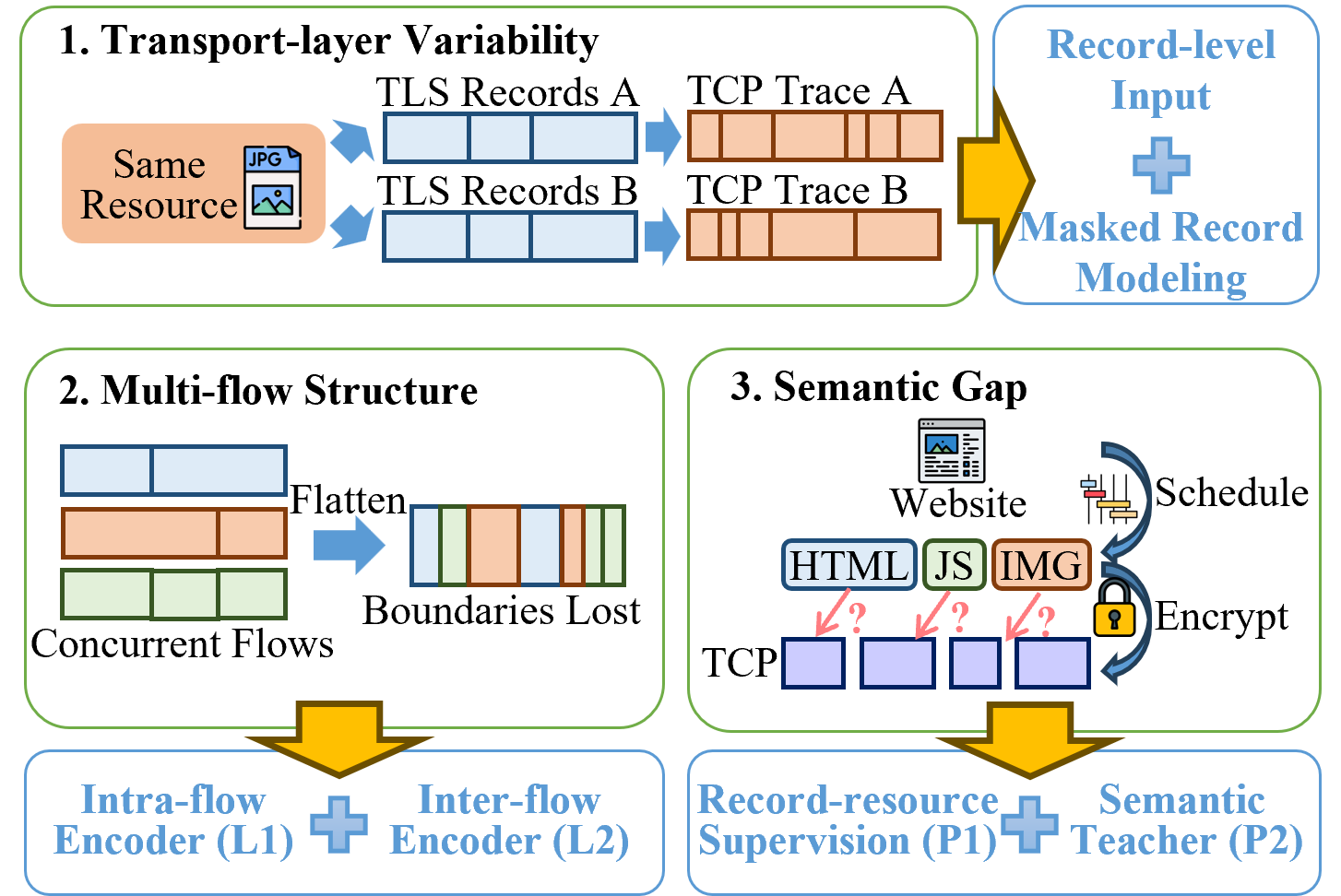}
    \caption{Challenges and motivation for CipherSight. Robust HTTPS WF requires reduced sensitivity to transport variability, explicit modeling of dependencies within and across flows, and supervision that connects encrypted records with web-resource structure. }
    \label{fig:motivation}
\end{figure}

To address these challenges, we propose CipherSight, a hierarchical transformer framework for robust HTTPS WF. Unlike existing packet-level representations that are highly affected by transport-layer dynamics, CipherSight models webpage loads at the TLS record level, which provides a more stable abstraction over TCP packets. To capture the hierarchical structure of modern HTTPS traffic, CipherSight introduces an intra-flow encoder using block-diagonal self-attention to learn dependencies among records within each flow and an inter-flow encoder that models interactions across concurrent flows for webpage-level representation learning. Furthermore, CipherSight exploits fine-grained resource structures as privileged supervision, aligning TLS record representations with resource boundaries and attributes to learn richer semantic representations beyond website-level labels. A privileged semantic teacher further transfers resource-level knowledge through LoRA fine-tuning. Privileged components are discarded at inference.
Figure~\ref{fig:motivation} summarizes the motivation and corresponding design principles.

Extensive experiments on more than 2,000 monitored websites show that CipherSight achieves 95.41\% closed-world accuracy and retains 92.99\% under temporal drift, exceeding the best-performing baseline by 13.03 percentage points (pp). It also maintains over 90\% accuracy in geographic drift settings, outperforming the best regional baselines by at least 8.52 pp, and achieves 97.64\% AUROC in open-world recognition. Ablations further confirm that privileged supervision provides its clearest gains under geographic drift.

\subsection{Contributions}

Our contributions are summarized as follows:
\begin{itemize}
    \item We design a structure-preserving representation for HTTPS WF that reduces dependence on TCP packetization, bringing it closer to higher-level semantics.
    \item We propose a hierarchical architecture that jointly models multiple TLS-record attributes and explicitly separates intra-flow dependencies from inter-flow interactions, which suppresses transport-specific artifacts without discarding the multi-flow structure of web pages.
    \item We introduce fine-grained privileged supervision based on web-resource semantics for HTTPS WF. CipherSight not only aligns TLS-record states with resource boundaries and attributes, but also transfers resource knowledge through privileged distillation during LoRA fine-tuning.

    \item We evaluate CipherSight on six HTTPS collections covering more than 2,000 websites under closed-world, temporal-drift, geographic-drift, and open-world settings. The results demonstrate gains in closed-world top-1 accuracy, temporal and geographic robustness, and open-world AUROC over representative baselines.
\end{itemize}

\section{Related Work}
\subsection{Website Fingerprinting}
Historically, HTTPS destinations could often be identified directly from hostnames exposed in plaintext DNS queries or the SNI field of TLS ClientHello~\citep{chai2019esni}. Visited websites can also be inferred from the sets and sequences of contacted destination IP addresses~\citep{hoang2021ipwf}. However, encrypted DNS and ECH increasingly conceal hostname signals, while CDN deployment, shared hosting, DNS-based load balancing, and IP address churn weaken direct domain-to-IP mappings. Consequently, reliable website identification increasingly depends on side-channel patterns that remain observable in encrypted traffic under realistic network environments.

Inferring visited websites from encrypted traffic side channels is commonly known as website fingerprinting (WF). WF has been studied most extensively in Tor, an anonymity network designed to conceal communication destinations~\citep{tor,tor_wf}. Tor-oriented methods commonly represent a page load as a flat sequence of observable packet or cell directions, bursts, and timings. Convolutional models such as AWF, DF, and VarCNN learn discriminative representations from directional sequences~\citep{awf,df,varcnn}, while TikTok explicitly incorporates timing information~\citep{tiktok}. Although effective on Tor traces, these flat sequence representations, when transferred to HTTPS traffic, do not explicitly preserve TLS-record boundaries or the concurrent multi-flow structure of modern page loads.

\subsection{Website Fingerprinting over HTTPS}
HTTPS-native WF methods increasingly move beyond flat TCP packet sequences. H\&W derives lightweight fingerprints from HTTP-version-dependent parallel loading, while ADU recovers application data unit length sequences before classification~\citep{hw,adu}. STC-WF models inter-flow spatio-temporal correlations using a graph neural network, and CTX-Aware exploits flow context in encrypted-proxy traffic~\citep{stcwf,ma2024flowcontext}. These studies demonstrate the benefits of higher-layer and flow-aware representations, but do not jointly encode multiple TLS-record attributes with explicit intra-flow and inter-flow dependencies in a unified hierarchy.

Semantic information has also been explored for improving WF generalization. STAR aligns encrypted traffic traces with crawl-time semantic profiles for zero-shot website retrieval~\citep{star}. Related preprints investigate resource-level semantic distillation and semantics-aware traffic augmentation~\citep{resaware,sata}. CipherSight differs by treating fine-grained alignment between TLS records and resource spans as privileged information under the LUPI paradigm~\citep{vapnik2009privileged}, combining record- and span-level supervision with privileged distillation in a hierarchical ciphertext encoder.

\subsection{Website Fingerprinting under Distribution Shifts}
High performance in closed-world settings cannot guarantee generalization to real-world conditions.
Prior Tor studies show that changes in collection time, network environment, and browsing context, together with unmonitored traffic, can induce distribution shifts that substantially degrade WF performance~\citep{tor_wf,cherubin2022online,jansen2023repositioning,yuan2024hswf,ares}. Temporal and geographic drift alter the input distribution over monitored classes, whereas open-world recognition additionally requires rejecting websites unseen during training~\citep{koh2021wilds,wang2023dgsurvey}. Existing robustness studies primarily focus on Tor or isolate a single source of mismatch. We therefore evaluate HTTPS WF under temporal drift, geographic drift, and open-world recognition to assess robustness across complementary deployment conditions.

\section{Method}
\begin{figure*}[t]
    \centering
    \includegraphics[width=1\textwidth]{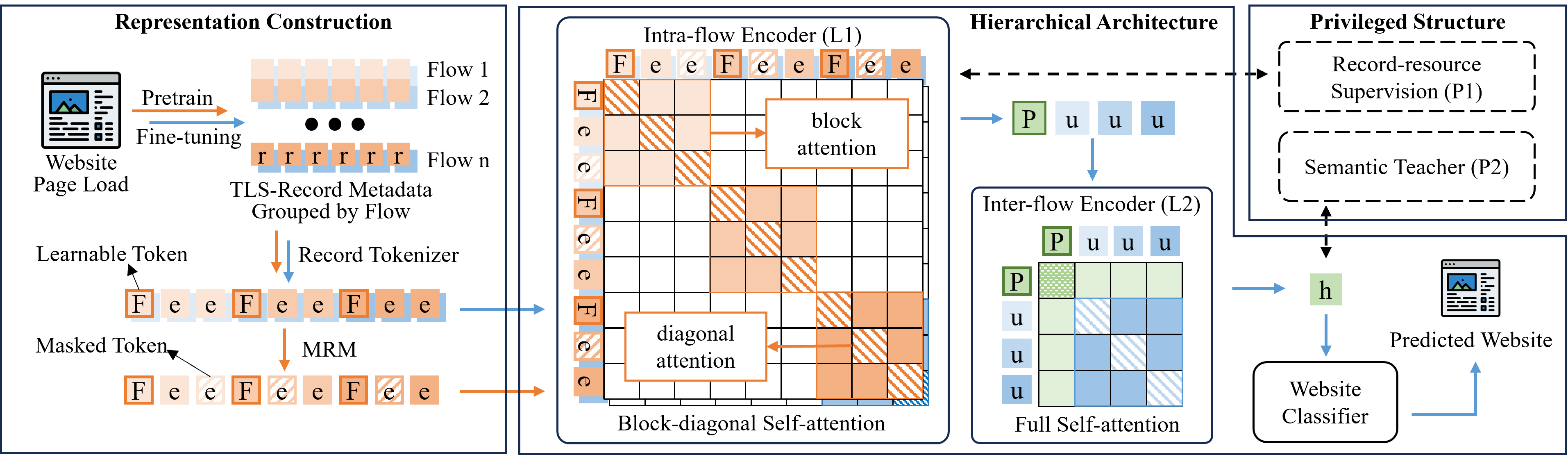}
    \caption{Overview of \textsc{CipherSight}. During representation construction, observable TLS records are grouped by flow and tokenized. Block-diagonal intra-flow attention (L1) encodes record tokens within each flow, while full inter-flow attention (L2) aggregates flow summaries for website classification. Training-time privileged structure provides record-resource supervision (P1) and semantic teacher distillation (P2).}
    \label{fig:architecture}
\end{figure*}
\subsection{Overview}

Figure~\ref{fig:architecture} presents CipherSight, whose representation construction, hierarchical architecture, and privileged structure are motivated by the three challenges identified in Figure~\ref{fig:motivation}. First, CipherSight represents traffic at the TLS-record level rather than the TCP-packet level, reducing sensitivity to transport-layer packetization. Ciphertext-observable record attributes are extracted and tokenized, while masked record modeling (MRM) guides pretraining. Second, record tokens are processed by an intra-flow encoder (L1) and an inter-flow encoder (L2), which separately model dependencies within flows and interactions across flows, reflecting the multi-flow organization of HTTPS page loads. Third, training-time privileged supervision connects ciphertext patterns with webpage structure. Record-resource supervision (P1) is used during pretraining and fine-tuning, while a semantic teacher (P2) is introduced during fine-tuning.

\subsection{Representation and Hierarchical Encoding}

\subsubsection{TLS-Record Representation and Tokenization.}
TLS records are protocol data units carried above TCP. After TCP reassembly, their boundaries and outer headers can be recovered without decryption~\citep{rfc8446,rfc9293}. A TLS record may span multiple TCP segments, while a single segment may contain bytes from multiple records. Consequently, record-level modeling is less directly coupled to TCP segmentation and retransmission than packet-level modeling, although it remains affected by changes in website content and TLS behavior.

In the representation construction phase, TLS records extracted from webpage-load trace $r_{i,j}$ are grouped into flows $f_i$. Categorical features are mapped to learnable embeddings, whereas numerical features combine logarithmic bucket embeddings with linear projections of their log-scaled values. The resulting feature vectors are summed and normalized to produce the record token $\mathbf{e}_{i,j}$.

\subsubsection{Intra-Flow Encoding (L1).}
Flattening all records into a single sequence can create direct dependencies between unrelated connections and make the representation sensitive to their collection-specific interleaving. L1 instead encodes records from the same flow to preserve TLS-flow boundaries. For every valid flow $f_i$, we prepend an instance of a shared learnable token $\mathrm{FCLS}_i$ to form
$(\mathrm{FCLS}_i,\mathbf{e}_{i,1},\ldots,\mathbf{e}_{i,n_i})$.
All flow sequences are concatenated for batched processing by a shared Transformer.

Let $\phi(p)$ denote the flow associated with token position $p$. L1 applies the attention bias
\begin{equation}
\label{eq:l1_mask}
M^{(1)}_{pq}=
\begin{cases}
0, & \phi(p)=\phi(q) \text{ and } p,q \text{ are valid},\\
-\infty, & \text{otherwise}.
\end{cases}
\end{equation}
The block-diagonal mask uses each TLS flow as a local context boundary and restricts self-attention to records from that flow. Copies of a shared learnable \texttt{[FCLS]} token aggregate within-flow information into flow summaries $\mathbf{u}_i$, while $\mathbf{h}_{i,j}$ denotes the contextual representation of each record. Rotary positions are reset within each flow, and cross-flow interaction is deferred to L2.

\subsubsection{Inter-Flow Encoding (L2).}
A modern website load typically involves connections from multiple domains, so a single flow cannot represent the complete webpage structure. CipherSight therefore prepends a learnable page token \texttt{[PCLS]} to the valid flow summaries and feeds
$(\mathrm{PCLS},\mathbf{u}_1,\ldots,\mathbf{u}_m)$
to the inter-flow encoder L2. Unlike L1, L2 applies full self-attention across valid flow representations.

Using flow summaries as an interface between L1 and L2 avoids all-to-all mixing between individual records from different connections while allowing webpage-level information to be combined across flows. The final hidden state of \texttt{[PCLS]} serves as the student page representation $\mathbf{h}_s$.
A website classifier consisting of a linear layer, GELU activation, dropout, and a final class projection produces the website logits $\mathbf{o}_s$.

\subsection{Pretraining with Record-Resource Supervision}

\begin{figure*}[t]
    \centering
    \includegraphics[width=1\linewidth]{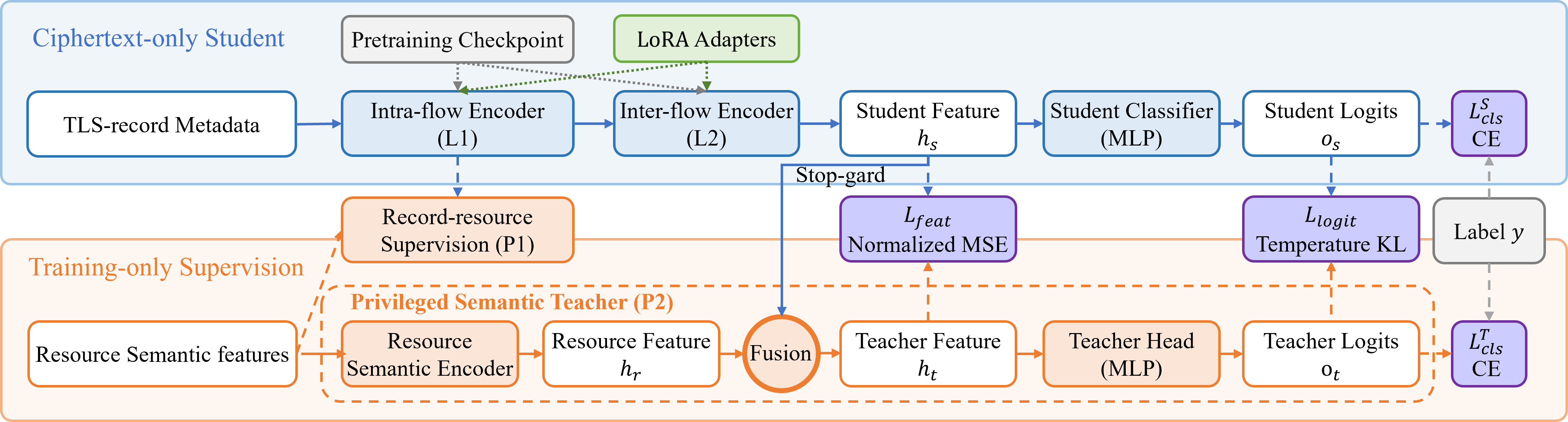}
    \caption{Fine-tuning architecture of CipherSight. The LoRA-adapted ciphertext student receives resource-structure supervision (P1), while a privileged semantic teacher (P2) transfers resource-level knowledge through feature and logit distillation, $L_{\mathrm{feat}}$ and $L_{\mathrm{logit}}$. The student and teacher are supervised by $L_{\mathrm{cls}}^{S}$ and $L_{\mathrm{cls}}^{T}$, respectively. Only the student is retained for inference.}
    \label{fig:finetune}
\end{figure*}

\subsubsection{Masked Record Modeling (MRM).}

Website labels do not directly supervise individual record representations. MRM replaces the complete embeddings of randomly selected valid records with a learnable \texttt{[MASK]} token and requires L1 to reconstruct their direction, length bucket, and outer TLS content type from the remaining within-flow context. The three cross-entropy losses are summed as $L_{\mathrm{MRM}}$, enabling contextual record modeling without resource annotations.

\subsubsection{Record-resource Supervision (P1).}

MRM captures ciphertext context but does not explicitly model the correspondence between TLS records and webpage resources. Prior work has demonstrated that resource semantics can improve the robustness of WF representations~\citep{resaware}. Motivated by this observation, we provided novel training-time alignments between website resources and TLS records in P1 to supervise L1 at both record and resource levels.

At the record level, P1 predicts begin/inside labels and inner MIME types over aligned records, while unaligned records are ignored by these losses. At the resource level, aligned record states are mean-pooled to predict response attributes and record counts. Algorithm~\ref{alg:p1_supervision} summarizes this two-level supervision and pooling procedure.

\begin{algorithm}[ht]
\caption{Record-Resource Supervision and Aligned-Record Pooling}
\label{alg:p1_supervision}
\textbf{Input}: Record states $\mathbf{H}$, record targets
$\mathbf{Y}_{\mathrm{BI}}$ and $\mathbf{Y}_{\mathrm{type}}$,
retained aligned spans $\{I_q\}$, and resource attributes
$\mathbf{Y}_{\mathrm{res}}$\\
\textbf{Output}: $L_{\mathrm{seg}},L_{\mathrm{attr}},L_{\mathrm{cnt}}$
\begin{algorithmic}[1]
\STATE $(\widehat{\mathbf{Y}}_{\mathrm{BI}},
        \widehat{\mathbf{Y}}_{\mathrm{type}})
        \leftarrow H_{\mathrm{rec}}(\mathbf{H})$.
\STATE Compute $L_{\mathrm{seg}}$ over aligned record positions.
\STATE Initialize $G\leftarrow\emptyset$ and $T\leftarrow\emptyset$.
\FOR{each resource $q$ with a nonempty retained span $I_q$}
    \STATE $\mathbf{g}_q\leftarrow
        |I_q|^{-1}\sum_{(i,j)\in I_q}\mathbf{h}_{i,j}$.
    \STATE Append $\mathbf{g}_q$ to $G$.
    \STATE Append $(y_q^{\mathrm{len}},y_q^{\mathrm{mime}},
        \log(1+|I_q|))$ to $T$.
\ENDFOR
\STATE $(\widehat{\mathbf{y}}^{\mathrm{len}},
        \widehat{\mathbf{y}}^{\mathrm{mime}},\widehat{\mathbf{c}})
        \leftarrow H_{\mathrm{res}}(G)$.
\STATE Compute $L_{\mathrm{attr}}$ and $L_{\mathrm{cnt}}$ using $T$.
\STATE \textbf{return} $L_{\mathrm{seg}},L_{\mathrm{attr}},L_{\mathrm{cnt}}$.
\end{algorithmic}
\end{algorithm}

\subsubsection{Pretraining Objective.}
The pretraining stage objective combines ciphertext-context reconstruction from MRM with resource-structure supervision (P1):
\begin{equation}
\label{eq:pretrain_loss}
\begin{aligned}
L_{\mathrm{P1}}
&=\lambda_{\mathrm{seg}}L_{\mathrm{seg}}
+\lambda_{\mathrm{attr}}L_{\mathrm{attr}}
+\lambda_{\mathrm{cnt}}L_{\mathrm{cnt}},\\
L_{\mathrm{pre}}
&=\lambda_{\mathrm{MRM}}L_{\mathrm{MRM}}+L_{\mathrm{P1}}.
\end{aligned}
\end{equation}
MRM learns dependencies among observable records, whereas P1 relates their hidden states to web-resource structure. The pretraining stage directly optimizes the tokenizer and L1, and L2 is introduced for website classification in the fine-tuning stage.

\subsection{Privileged Semantic Fine-Tuning}

\subsubsection{LoRA Adaptation.}
Pretraining establishes record-level representations, which should not be overwritten by unrestricted task adaptation. We therefore freeze the pretrained backbone and introduce low-rank updates
$\widetilde{W}=W+(\alpha/r)BA$
into the query, key, value, and output projections of every L1 and L2 attention layer. The LoRA matrices, student classifier, P1 heads, and semantic teacher remain trainable. This allows L2 to learn webpage-level aggregation while limiting changes to the pretraining representation.

\subsubsection{Privileged Semantic Teacher (P2).}

Website-label fine-tuning alone may emphasize environment-specific correlations and weaken the semantic structure learned during pretraining. As shown in Figure~\ref{fig:finetune}, privileged semantic teacher (P2) supplies website-level semantic guidance using resource attributes available only during training.

P2 first constructs resource tokens from resource semantic features.
Then a resource semantic encoder summarizes the resource token sequence as $\mathbf{h}_r$. The teacher fuses $\mathbf{h}_r$ with a stop-gradient copy of the student feature $\mathbf{h}_s$ to produce $\mathbf{h}_t$, and an MLP produces teacher logits $\mathbf{o}_t$.
The teacher is trained with website-label cross-entropy and uses $L_{\mathrm{feat}}$ and $L_{\mathrm{logit}}$ to distill resource semantics into the student.
Formally, let
$\bar{\mathbf{h}}_x=N(\mathbf{h}_x)$ and
$\mathbf{p}_x=\operatorname{softmax}(\mathbf{o}_x/T)$ for
$x\in\{s,t\}$. The distillation losses are
\begin{equation}
\label{eq:distill_losses}
\begin{aligned}
L_{\mathrm{feat}}
  &= \mathrm{MSE}\bigl(\bar{\mathbf{h}}_s,
     \operatorname{sg}[\bar{\mathbf{h}}_t]\bigr),\\
L_{\mathrm{logit}}
  &= T^2\mathrm{KL}\bigl(
     \operatorname{sg}[\mathbf{p}_t]\,\|\,\mathbf{p}_s\bigr).
\end{aligned}
\end{equation}
Here, $N(\cdot)$ denotes L2 normalization, $T$ is the distillation
temperature, and $\operatorname{sg}$ stops gradients.

\begin{table*}[t]
\centering
\begingroup
\fontsize{9pt}{10pt}\selectfont
\begin{tabular}{llccccc}
\toprule
Model & Origin & Accuracy & Precision & Recall & Macro-F1 & Top-5 \\
\midrule
VarCNN & Tor & 92.16\% & 92.63\% & 92.02\% & 91.91\% & 97.27\% \\
ARES   & Tor & 91.22\% & 91.65\% & 91.07\% & 90.85\% & 97.11\% \\
RF     & Tor & 91.09\% & 91.27\% & 90.84\% & 90.63\% & 97.16\% \\
DF     & Tor & 89.85\% & 90.29\% & 89.56\% & 89.36\% & 96.61\% \\
TikTok & Tor & 89.32\% & 89.69\% & 89.04\% & 88.77\% & 96.50\% \\
TF     & Tor & 86.94\% & 87.37\% & 86.62\% & 86.41\% & -- \\
AWF    & Tor & 62.66\% & 66.33\% & 62.40\% & 61.95\% & 81.59\% \\
\midrule
H\&W   & HTTPS & 86.45\% & 86.67\% & 86.23\% & 85.76\% & 94.31\% \\
STC-WF & HTTPS & 77.75\% & 80.00\% & 77.66\% & 77.34\% & 90.65\% \\
CTX-Aware & HTTPS & \phantom{0}8.18\% & \phantom{0}7.00\% & \phantom{0}8.14\% & \phantom{0}5.88\% & -- \\
\midrule
CipherSight & HTTPS
& \textbf{95.41$\pm$0.11\%}
& \textbf{94.85$\pm$0.14\%}
& \textbf{95.00$\pm$0.13\%}
& \textbf{94.67$\pm$0.14\%}
& \textbf{97.71$\pm$0.02\%} \\
\bottomrule
\end{tabular}
\endgroup
\caption{Closed-world performance comparison on 2,008 website classes. CipherSight results are reported as the mean $\pm$ standard deviation over five seeds.}
\label{tab:closeworld}
\end{table*}

\subsubsection{Fine-Tuning Objective and Inference.}

Let $L_{\mathrm{cls}}^{S}=\mathrm{CE}(\mathbf{o}_s,y)$ and
$L_{\mathrm{cls}}^{T}=\mathrm{CE}(\mathbf{o}_t,y)$ denote the student and teacher classification losses. The fine-tuning stage minimizes
\begin{equation}
\label{eq:finetune_loss}
\begin{aligned}
L_{\mathrm{ft}}={}&
\lambda_{\mathrm{cls}}L_{\mathrm{cls}}^{S}
+\lambda_{\mathrm{struct}}
(L_{\mathrm{seg}}+L_{\mathrm{attr}}+L_{\mathrm{cnt}})
+L_{\mathrm{cls}}^{T}\\
&+\lambda_{\mathrm{logit}}L_{\mathrm{logit}}
+\lambda_{\mathrm{feat}}L_{\mathrm{feat}}.
\end{aligned}
\end{equation}
Continuing P1 helps preserve record-resource structure during adaptation, while P2 transfers webpage-level semantics to the ciphertext student. MRM is used only in pretraining. At inference, CipherSight uses only TLS-record metadata, the L1 and L2 encoders, and the website classifier. The MRM, P1, P2, and all resource annotations are removed.

\section{Evaluation}
\subsection{Datasets and Experimental Setup}

\textbf{Collection and Construction.}
Each automated webpage visit produces a PCAP trace and a corresponding SSL key log. We reconstruct TCP flows and derive the model inputs from ciphertext-observable TLS metadata, including record direction, length, outer content type, timing, and flow membership, without decrypting application payloads. For dataset annotation only, we use the SSL key log with standards-compliant TLS and HTTP parsing to recover the record-resource alignments and resource attributes required by P1 and P2.

\textbf{Datasets.}
We evaluate six collections spanning four dates and three regions. Closed-world evaluation uses an 8:2 split of us0304, yielding 64,386 training and 15,848 test traces from 2,008 classes. Temporal drift trains on 80\% of \texttt{us0304} and tests \texttt{us0320}, collected 16 days later, over 1,976 shared classes. Geographic drift trains on 80\% of \texttt{us0409} and tests same-day \texttt{fr0409} and \texttt{sg0409} over 1,936 shared classes. The held-out 20\% of \texttt{us0409} traces measure performance degradation. Open-world evaluation reuses the closed-world split and adds 213,099 traces from disjoint websites.

\textbf{Metrics.}
We report top-1 accuracy, macro-F1, and top-5 accuracy where supported. Open-world confidence is maximum softmax probability. With monitored traffic as the positive class, AUROC and AUPR measure known-unknown discrimination, while OSCR jointly reflects monitored classification and unknown rejection across thresholds.

\textbf{Baselines.}
We compare Tor-oriented methods \citep{awf,df,varcnn,tiktok,tf,rf,ares} and HTTPS-native methods \citep{hw,stcwf,ma2024flowcontext}. Tor baselines use 5,000-packet sequences, truncated or zero-padded, with direction, length, and timing when applicable. Missing preprocessing or open-world evaluation is implemented from reported definitions without altering model architecture.

Neural baselines train for 100 epochs, except STC-WF for 300. To favor baselines, we select closed-world checkpoints by peak test accuracy and drift checkpoints on the fixed source-domain 20\% holdout before evaluation on complete target sets. Open-world evaluation reuses those checkpoints. CipherSight uses its final 20,000-step checkpoint.

\textbf{Implementation.}
CipherSight uses PyTorch, 8 NVIDIA GPUs, and BF16 precision. Both stages run for 20,000 steps.
CipherSight uses seeds 42--46, and the baselines use seed 42 unless stated otherwise.

\subsection{Closed-World Evaluation}

As shown in Table~\ref{tab:closeworld}, CipherSight achieves the best closed-world performance across 2,008 website classes, reaching 95.41\% accuracy and 94.67\% macro-F1. It outperforms VarCNN, the strongest baseline, by 3.25 and 2.76 percentage points (pp) on these metrics, respectively. The standard deviations remain below 0.15 pp across five seeds, indicating stable performance across runs. By comparison, the top-5 improvement over VarCNN is only 0.44 pp. This contrast indicates that CipherSight primarily improves first-choice discrimination rather than merely retaining the correct website among several candidates.

STC-WF warrants a qualified interpretation. A separate run with seed 52 on the same split reaches 84.78\% accuracy, while it achieves 92.25\% on the \texttt{us0409} source-domain holdout under the geographic-drift protocol. These results indicate sensitivity to training randomness rather than uniformly weak performance.

CTX-Aware achieves only 8.18\% accuracy under our adaptation. Its full configuration required more than 128~GB of memory, so we restricted each webpage trace to six flows and used a smaller random forest.

\subsection{Temporal Drift}

\begin{table}[htbp]
    \centering
    \begingroup
    \fontsize{9pt}{10pt}\selectfont
    \begin{tabular}{lll}
    \toprule
        Model & Accuracy (\%) & Macro-F1 (\%) \\
    \midrule
    VarCNN & 67.81 (-24.35) & 67.30 (-24.61) \\
    ARES   & 72.84 (-18.38) & 71.92 (-18.93) \\
    RF     & 66.07 (-25.02) & 65.82 (-24.81) \\
    DF     & 67.92 (-21.93) & 66.61 (-22.75) \\
    TikTok & 66.03 (-23.29) & 64.74 (-24.03) \\
    TF     & 67.75 (-19.19) & 66.49 (-19.92) \\
    AWF    & 29.10 (-33.56) & 29.07 (-32.88) \\
    \midrule
    H\&W   & 79.96 (-\phantom{0}6.49)
           & 77.92 (-\phantom{0}7.84) \\
    STC-WF & 73.93 (-\phantom{0}3.82)
           & 72.50 (-\phantom{0}4.83) \\
    \midrule
    CipherSight & \textbf{92.99 (-\phantom{0}2.42)}
                & \textbf{92.07 (-\phantom{0}2.60)} \\
    \bottomrule
    \end{tabular}
    \endgroup
    \caption{Temporal-drift performance from \texttt{us0304} to \texttt{us0320} with a 16-day collection gap. Values in parentheses denote descriptive percentage-point differences from each model's closed-world result on \texttt{us0304}.}
    \label{tab:time-drift}
\end{table}

\begin{table*}[t]
\centering
\begingroup
\fontsize{9pt}{10pt}\selectfont
\begin{tabular}{lcccccc}
\toprule
       & \multicolumn{3}{c}{Singapore} & \multicolumn{3}{c}{France} \\
       \cmidrule(lr){2-4} \cmidrule(lr){5-7}
Model & Accuracy (\%) & Macro-F1 (\%) & Top-5 (\%)
      & Accuracy (\%) & Macro-F1 (\%) & Top-5 (\%) \\
\midrule
VarCNN
& 67.53 (-24.98) & 66.20 (-26.24) & 83.07 (-14.30)
& 73.12 (-19.39) & 70.70 (-21.74) & 85.05 (-12.32) \\

ARES
& 79.53 (-13.16) & 78.15 (-14.48) & 89.97 (-\phantom{0}7.23)
& 81.08 (-11.61) & 79.11 (-13.52) & 90.58 (-\phantom{0}6.62) \\

RF
& 57.69 (-35.12) & 57.44 (-35.19) & 73.25 (-24.10)
& 82.07 (-10.74) & 80.19 (-12.44) & 91.03 (-\phantom{0}6.32) \\

DF
& 74.95 (-17.39) & 72.83 (-19.26) & 87.20 (-10.10)
& 76.48 (-15.86) & 73.24 (-18.85) & 86.41 (-10.89) \\

TikTok
& 70.00 (-21.98) & 67.80 (-23.94) & 84.26 (-13.14)
& 74.51 (-17.47) & 71.24 (-20.50) & 85.08 (-12.32) \\

TF
& 74.46 (-16.56) & 72.44 (-18.43) & --
& 75.67 (-15.35) & 72.70 (-18.17) & -- \\

AWF
& 32.09 (-36.74) & 31.93 (-36.70) & 54.92 (-30.77)
& 45.99 (-22.84) & 45.18 (-23.45) & 67.52 (-18.17) \\
\midrule
H\&W
& 75.94 (-14.82) & 72.87 (-17.35) & 84.52 (-11.82)
& 67.56 (-23.20) & 62.57 (-27.65) & 76.56 (-19.78) \\

STC-WF
& 60.60 (-31.65) & 57.52 (-34.32) & 75.12 (-22.60)
& 54.38 (-37.87) & 49.74 (-42.09) & 68.96 (-28.76) \\
\midrule
CipherSight
& \textbf{91.09 (-\phantom{0}4.41)} & \textbf{89.99 (-\phantom{0}5.21)} & \textbf{95.20 (-\phantom{0}2.50)}
& \textbf{90.59 (-\phantom{0}4.91)} & \textbf{89.19 (-\phantom{0}6.01)} & \textbf{95.08 (-\phantom{0}2.63)} \\
\bottomrule
\end{tabular}
\endgroup
\caption{Geographic-drift performance from \texttt{us0409} to Singapore and France. Values in parentheses denote descriptive percentage-point differences from each model's \texttt{us0409} source-domain holdout. CipherSight values are means over five seeds.}
\label{tab:geo-drift}
\end{table*}

Table~\ref{tab:time-drift} reports performance under a temporal distribution shift caused by a 16-day gap in data collection. H\&W is the strongest baseline with 79.96\% accuracy, while STC-WF reaches 73.93\% and slightly outperforms ARES, the strongest Tor-oriented baseline. Although both HTTPS-specific methods trail the leading Tor-oriented baselines in the closed-world setting, their ranking reverses under temporal drift. This result indicates that in-distribution closed-world accuracy is not a reliable proxy for robustness to temporal drift.

CipherSight achieves 92.99\% accuracy and 92.07\% macro-F1, outperforming H\&W by 13.03 and 14.15 percentage points (pp), respectively. Its descriptive differences from the separate closed-world result are only $-2.42$ and $-2.60$ pp, smaller than those of all baselines. Its accuracy margin over H\&W also widens from 8.96 pp in the closed-world setting to 13.03 pp under temporal drift. Although these differences are not paired source-to-target drops, the pattern is consistent with CipherSight learning representations that remain informative as website traffic evolves.

\subsection{Geographic Drift}

Table~\ref{tab:geo-drift} evaluates geographic drift by training on \texttt{us0409} and testing on the same-day \texttt{sg0409} and \texttt{fr0409} collections. This design reduces temporal confounding and helps isolates regional variation.

Baseline rankings vary by region. ARES is the strongest baseline in Singapore with 79.53\% accuracy, whereas RF leads in France with 82.07\%. RF differs by 24.38 percentage points (pp) between regions, showing that performance in one region does not reliably predict another.

CipherSight achieves the highest accuracy, macro-F1, and top-5 accuracy in both regions, reaching 91.09\% accuracy in Singapore and 90.59\% in France. It outperforms the strongest regional baselines by 11.56 and 8.52 pp, respectively. Its accuracy differs by only 0.50 pp between regions, and it exhibits the smallest degradation under geographic drift across all reported metrics. These results suggest that CipherSight retains stable record- and flow-level evidence across regions.

\subsection{Ablation Study}

\begin{table}[ht]
\centering
\begingroup
\fontsize{9pt}{10pt}\selectfont
\begin{tabular}{lccc}
\toprule
Test & No-Priv (\%) & P1 Pre. (\%) & Full Priv. (\%) \\
\midrule
us0320 & 92.69 / 91.73 & 92.99 / 92.07 & 92.99 / 92.07 \\
us0409 & 90.46 / 89.02 & 90.93 / 89.53 & 91.11 / 89.73 \\
\midrule
fr0409 & 89.04 / 87.37 & 90.15 / 88.73 & 90.59 / 89.19 \\
sg0409 & 89.73 / 88.42 & 90.59 / 89.46 & 91.09 / 89.99 \\
\bottomrule
\end{tabular}
\endgroup
\caption{Ablation of privileged supervision under temporal and geographic drift. No-Priv. excludes P1 and P2, P1 Pre. applies P1 only during pretraining, and Full Priv. applies P1 in both stages and introduces P2 during LoRA fine-tuning. Cells report mean accuracy/macro-F1 (\%) over five seeds.}
\label{tab:ablation-lora-priv}
\end{table}

Table~\ref{tab:ablation-lora-priv} reports the contribution of privileged supervision under distribution shifts. Under temporal drift, P1 Pre. improves accuracy over No-Priv. by 0.30 and 0.47 percentage points (pp) on \texttt{us0320} and \texttt{us0409}, while Full Priv. yields gains of 0.30 and 0.65 pp. This result shows that P1 pretraining provides most of the temporal benefit.

The contribution is more pronounced under geographic drift. P1 Pre. improves accuracy by 0.86 pp on \texttt{sg0409} and 1.11 pp on \texttt{fr0409}, while Full Priv. increases these gains to 1.37 and 1.55 pp, with corresponding macro-F1 gains of 1.58 and 1.83 pp. Although modest in absolute terms, these gains are approximately 31\% as large as CipherSight's overall accuracy degradation under geographic drift, with a similar ratio of about 30\% for macro-F1. Overall, the ablation demonstrates the effectiveness of training-time privileged supervision, particularly in improving geographic robustness.

\subsection{Open-World Evaluation}

\begin{table}[htbp]
    \centering
    \begingroup
    \fontsize{9pt}{10pt}\selectfont
    \begin{tabular}{lccc}
    \toprule
        Model & AUROC (\%) & AUPR (\%) & OSCR (\%) \\
    \midrule
        VarCNN & 89.56 & 60.50 & 86.22 \\
        ARES   & 89.93 & 67.27 & 86.05 \\
        RF     & 93.04 & 72.03 & 87.61 \\
        DF     & 87.06 & 63.46 & 83.72 \\
        TikTok & 86.21 & 61.49 & 82.77 \\
        TF     & 82.75 & 26.95 & 77.15 \\
        AWF    & 70.11 & 27.48 & 52.40 \\
    \midrule
        H\&W   & 57.23 &  \phantom{0}7.39 & 49.42 \\
        STC-WF & 77.92 & 35.69 & 68.67 \\
    \midrule
        CipherSight & \textbf{97.64} & \textbf{88.54} & \textbf{94.81} \\
    \bottomrule
    \end{tabular}
    \endgroup
    \caption{Open-world evaluation results using maximum softmax probability (MSP) as the confidence score.}
    \label{tab:open-world}
\end{table}

As shown in Table~\ref{tab:open-world}, CipherSight consistently outperforms all baselines, achieving 97.64\% AUROC, 88.54\% AUPR, and 94.81\% OSCR. Relative to RF, the strongest baseline, it yields absolute gains of 4.60, 16.51, and 7.20 percentage points, respectively. The substantial improvements in AUPR and OSCR demonstrate more effective rejection of unmonitored samples in the highly imbalanced open-world setting, without compromising monitored-site classification.

\section{Limitations}
More refined data preprocessing methods and sampling-window strategies may further improve model performance~\citep{peng2025}. CipherSight could be extended to a range of scenarios, including longer-term temporal drift, larger numbers of clients, and HTTP/3 traffic. In addition, privileged supervision requires SSL keys, limiting packet-only reuse, although inference remains ciphertext-only.

\section{Conclusion}
We presented CipherSight, a hierarchical Transformer for robust HTTPS WF under distribution shifts. CipherSight models ciphertext-observable TLS records within and across flows. It uses record-resource supervision and privileged semantic distillation during training while preserving ciphertext-only inference. Across closed-world, temporal, geographic, and open-world evaluations, CipherSight consistently outperformed the baselines. Ablations showed gains from privileged supervision, particularly under geographic drift. Together, these results highlight the value of stable TLS-record representations, explicit inter-flow modeling, and training-time resource semantics for robust encrypted-traffic learning under the evaluated conditions.

\bibliography{aaai2027}

\end{document}